\documentclass[smallextended]{svjour3}
\smartqed  

\usepackage{graphicx}
\usepackage{multirow}
\usepackage{amsmath}
\usepackage{amssymb}
\usepackage{booktabs}
\usepackage{xspace}
\usepackage{xcolor}
\usepackage{url}
\usepackage{natbib}
\usepackage{tikz}
\usepackage{framed}
\usepackage{enumitem}
\newenvironment{rqanswer}[1]%
  {\begin{framed}\noindent\textbf{Takeaway #1:} }%
  {\end{framed}}
\usetikzlibrary{arrows.meta, positioning, decorations.pathreplacing, calc}
\usepackage[hidelinks]{hyperref}

\newcommand{\aidev}{AIDev\xspace}

\newcommand{\rqoneq}{How often are the merged agent PRs followed by a fix compared to the merged human PRs in the same repositories?}
\newcommand{\rqtwoq}{Who authors the verified fix PRs of the merged agent PRs?}
\newcommand{\rqthreeq}{To what extent are the commits pushed to the verified fix PRs authored by agents?}
\newcommand{\rqfourq}{How do merge-time signals differ between merged agent PRs that are followed by a fix and those that are not?}

\newcommand{\rqonea}{22.9\,\% of merged agent PRs are followed by a candidate fix and 4.5\,\% by a verified fix within 30 days.
In the same repositories and the same observation window, agent merges are fixed at  1.41 times the human odds (candidate) and 1.62 times (verified), and the fixes arrive early: half of the 30-day incidence accumulates within the first week.}
\newcommand{\rqtwoa}{69.6\,\% of the verified fixes on agent merges come from the same agent, while 89\,\% of the verified fixes on human merges also come from humans.
The finding shows that fixing work stays within the authoring population on both sides.}
\newcommand{\rqthreea}{The commits added to verified-fix agent PRs between first push and merge are, on average, 87.4\,\% agent-authored per PR, and 76.4\,\% of those PRs are agent-authored throughout all commits.}
\newcommand{\rqfoura}{Overall, merge-time signal distributions of agent PRs on both groups are nearly identical (all effects negligible to small). 
However, within repositories, the number of commits to complete a PR could mark a follow-up fix.
Specifically, a merge with ten times the commits carries $6.1$ times the odds of a verified fix.

}

\journalname{Empirical Software Engineering}

\begin{document}

\title{Who Finishes the Job? A Study of Follow-Up Fixes and Commit Authorship on AI Coding Agent Pull Requests}
\titlerunning{Who Finishes the Job on AI Coding Agent PRs?}

\author{Wannita Takerngsaksiri \and Nhat Duong  \and Scott Barnett}


\authorrunning{W. Takerngsaksiri et al.}

\institute{W. Takerngsaksiri \and N. Duong \and S. Barnett \at Applied AI Initiatives (A2I2), Deakin University, Melbourne, Australia \\
              \email{wannita.takerngsaksiri@deakin.edu.au}
}

\date{Received: date / Accepted: date}

\maketitle

\begin{abstract}
AI coding agents now author a large share of pull requests (PRs) merged into popular open-source projects.
A merged agent PR is usually considered finished work; yet, prior studies have reported issues in agent code after the merge (e.g., code smells and static-analysis issues).
However, little is known about how often a merged agent PR is fixed afterward, and who actually authors the fixing.
In this paper, we follow 6{,}774 merged agent PRs across five AI coding agents (OpenAI Codex, GitHub Copilot, Devin, Cursor, and Claude Code) from the \aidev{}-pop dataset (open-source repositories with at least 500 stars) into their follow-up fixes, against a baseline of 5{,}044 contemporaneous human PRs from the same repositories.
We link each merge to its candidate fixes, verify every candidate with human annotators and an LLM judge that matches human-level agreement (binary Cohen's $\kappa=0.78$ against a human--human $\kappa=0.77$, Direct-fix precision 90\,\%), and attribute the fixing work at the PR and the commit level.
Our findings show that (1) merged agent PRs attract verified fixes at 1.62 times the odds of merged human PRs in the same repositories over the same period of time; 
(2) 69.6\,\% of verified fixes in agent merges come from the same agent
; and 
(3) 76.4\,\% of the verified fix PRs are agent-authored throughout all commits.
These results show that agents currently largely finish their own job, but their merges still require fixing more often than human merges.

\end{abstract}

\section{Introduction}
\label{sec:introduction}


AI coding agents (e.g., Codex~\citep{openai2026codex}, Devin~\citep{devin2026devin}, Copilot~\citep{github2026copilot}, Cursor~\citep{cursor2026cursor}, and Claude Code~\citep{anthropic2026claude}) have revolutionised software engineering at a remarkable speed, transforming from assisting as auto-completion to taking authorship.
Agents now open, revise and merge pull requests (PRs) in their own workflows, on the scale of hundreds of thousands of repositories in both open source and industry~\citep{li2025aidev, robbes2026agentic, takerngsaksiri2025human}.

However, the review an agent PR passes before merging is quick and light: $61.4\%$ of agent PRs do not receive recorded human review~\citep{duma2026reviews}, and agent PRs complete within minutes to hours~\citep{watanabe2026communicate}.
Previous studies have also reported problems in agent code after the merge (e.g., static-analysis issues that persist~\citep{liu2026debt}, and later modifications that are corrective rather than adaptive~\citep{rahman2026survive, khemissi2026referenced}).
As a result, when a merged change contains an issue, the consequence is a \emph{follow-up fix}: a later PR that revisits and corrects the merged PR.
Understanding this follow-up fix caused by agents can help developers plan their tasks and how agent contributions should be evaluated. 
Nevertheless, little is known about these follow-up fixes on merged agent PRs.
Specifically, how \emph{often} merged agent PRs create follow-up fixes, \emph{who} comes back to fix them, and \emph{how} merged PRs with and without a follow-up fix differ have not been measured.

\textit{To address this gap,} we follow 6,774 merged agent PRs across 891 repositories and five AI coding agents in \aidev{}-pop dataset into their follow-up fixes.
We analyse \emph{how often} follow-up fixes occur, 
\emph{who} authors those fixes, and \emph{how} PRs with and without a follow-up fix differ, against a contemporaneous baseline of 5,044 merged human PRs from the same repositories.
We link each merge to its candidate follow-up fixes and verify every candidate with human annotators and a validated LLM judge.
Then, we classify who opened each fixing PR and who authored each of its commits.
Finally, we address the following four research questions:
\begin{description}
  \item[\textbf{RQ1.}] \textbf{\rqoneq}
  
  \textbf{Result.} \rqonea
  
  
  \item[\textbf{RQ2.}] \textbf{\rqtwoq}

  \textbf{Result.} \rqtwoa
  
  \item[\textbf{RQ3.}] \textbf{\rqthreeq}

  \textbf{Result.} \rqthreea
  
  \item[\textbf{RQ4.}] \textbf{\rqfourq}

  \textbf{Result.} \rqfoura

\end{description}

This paper makes the following contributions:
\begin{enumerate}
  \item We conduct an empirical study on the authorship of follow-up fixes of merged agent PRs at scale, evaluating against a human baseline in the same repositories over the same period of time.
  \item We perform a human validation to verify candidate fixes in agent PRs and an LLM judge that matches human-level agreement (Cohen's $\kappa=0.78$, Direct-fix precision 90\,\%) to ensure the relevance throughout the dataset.
  \item We perform a comprehensive analysis of the differences in merge-time signals between agent PRs with and without a follow-up fix. 
    \item Our replication package is released at \url{https://github.com/wannita901/pr-fix-authorship}.
\end{enumerate}

The remainder of the paper proceeds as follows: Section~\ref{sec:related-work}
positions our paper with other related works, 
Section~\ref{sec:methodology} presents the study design, 
Section~\ref{sec:results} reports each RQ's approach and results,
Section~\ref{sec:discussion} discusses our findings,
Section~\ref{sec:threats} reports
threats to validity, and Section~\ref{sec:conclusion} concludes our study.

\section{Background and Related Work}
\label{sec:related-work}


\subsection{AI Coding Agents in Software Development}
\label{sec:related-work:agents}

AI coding agents (e.g., Codex~\citep{openai2026codex}, Devin~\citep{devin2026devin}, Copilot~\citep{github2026copilot}, Cursor~\citep{cursor2026cursor}, and Claude
Code~\citep{anthropic2026claude}) have become routine contributors to open-source projects~\citep{robbes2026agentic, li2025aiteammates_se3} and industry~\citep{takerngsaksiri2025human}.
\cite{robbes2026agentic} studies more than 120,000 repositories and estimates the adoption rate of 22.20\,\%–28.66\,\% in February 2026.
\cite{li2025aiteammates_se3} describe this shift as the move from \emph{SE 2.0} (human-authored,
agent-assisted) to \emph{SE 3.0} (agent-authored, human-reviewed),
and release \aidev{}~\citep{li2025aiteammates_se3}, the dataset on which we build our analysis.
Nearly a million agent pull requests are distributed in more than a hundred thousand popular repositories open sources on GitHub in 2025.
On the other hand, \cite{takerngsaksiri2025human} conducted a study of the AI coding agent with human-in-the-loop, showing how agents were adopted at Atlassian, piloting over 600 real-world JIRA issues in production in a month.

Therefore, while the adoption rate of AI coding agents is increasing rapidly, this highlights the importance of our study in identifying what is left behind after an agent PR merges into production and who actually follow up to fix those issues~\textit{operationally}. 
%

\subsection{The Study of AI Coding Agent's Pull Requests}
\label{sec:related-work:prior-measurement}

Pull requests of AI coding agents have been a popular research topic in recent years.
\cite{li2025aidev} present AIDev dataset, which is a large-scale collection of agent commits and pull requests crawled from GitHub in 2025.
Subsequent works have grown rapidly around it in various angles: from studying at \emph{first-push characteristics}~\citep{ogenrwot2026modify, watanabe2026communicate, gong2026inconsistency}, \emph{review process}~\citep{duma2026reviews, haider2026themes}, \emph{merge outcomes}~\citep{peralta2026agentic, alam2026unmerged}, to \emph{after the merge}~\citep{liu2026debt, rahman2026survive, khemissi2026referenced}.


Starting at the \textit{first push} of an agent PR, previous studies analyse the characteristics of agent's signals compared to human PRs.
The studies found that agent PRs are smaller and more localised than human PRs~\citep{ogenrwot2026modify}, differ in description style~\citep{watanabe2026communicate}, and occasionally misdescribe their own diff~\citep{gong2026inconsistency}.

Next, during the \emph{code review process}, there are studies on the characteristics of review comments on AI agent PRs.
\cite{duma2026reviews} found that $61.4\,\%$ of agent PR do not receive human review records, while \cite{haider2026themes} found that the most common themes of comments on agent PRs are functional correctness.



Then, there are the \emph{merge outcomes} studies in which the merged and unmerged agent PRs have been analysed to find differences.
\cite{peralta2026agentic} found that unmerged outcomes substantially overstate agent error with only 35.7\,\% of rejected PRs reflected clear agent failures.
Additionally, \cite{alam2026unmerged} indicate test case failures and prior resolution of the same issues by other PRs are the most common causes of an unmerged PR.


Last but not least, there are studies that involve activities \emph{past the merge}, which our paper also falls into this category.
\cite{liu2026debt} found that $22.7\%$ of the static-analysis issues introduced by agents still survive to the latest version of the repositories at the time.
\cite{rahman2026survive} conducted survival analysis on agent code and found that agent code is modified \emph{less} often than human code, though its modifications are corrective more often than adaptive, unlike human code.
Similar to previous findings, \cite{khemissi2026referenced} mined the follow-up PRs that \emph{explicitly reference} an agent PR and found that more than half of agent-to-agent references are for corrective.


Different to prior works, we mine the \emph{follow-up fixes} of merged agent PRs, and verify the fix candidates with human validation and an LLM judge that aligns with human agreement.
Our goal is to study how often agent PRs receive a follow-up fix compared to a human baseline under the same conditions.

\subsection{Code Ownership in the Age of AI Coding Agents}
\label{sec:related-work:ownership}

Code ownership research traditionally links who authors a component to its quality.
\cite{bird2011ownership} find that the number of \emph{minor} contributors to a Windows binary correlates with its failures more strongly than size, churn, or complexity.
The study indicates that changes being routed to the component's owner rather than made by outsiders could minimise defects.
On the other hand, in open-source projects, the relationship is more nuanced.
Although the correlation between ownership and software quality is confirmed, ownership metrics add little beyond other measures~\citep{foucault2015ownership}.

In the age of AI coding agents, the question of code ownership has reopened.
Developers' felt ownership of code declines as agent autonomy increases~\citep{seo2026whosecode}, while the agents' terms of service consistently assign legal responsibility for the generated code to the human user~\citep{treude2026accountable}.
\cite{sawada2026maintenance} randomly tracked 508 agent-generated files and find that humans author 83.2\,\% of all commits, indicating a large human contribution over agent code. 
However, little is known about follow-up fix after a PR merges and who actually authors them.
Before the agents era, studies of self-fixing report that developers fix roughly half of their own defects at individual granularity: 55.7\,\% of single-statement bugs~\citep{zhu2021meaculpa}, 46\,\% of 8{,}033 fuzzer-detected bugs~\citep{keller2023ossfuzz}, and about half of fixed technical-debt issues~\citep{tan2022selffixed}, with a sense of responsibility for one's own code as the main reported motivation~\citep{tan2021selffixsurvey}.

Different to prior work, we measure the \emph{authorship} of a \emph{follow-up fix} after the agent PRs have been merged and whether the author class that merged a PR is the one that returns to fix it.

%


\section{Study Design}
\label{sec:methodology}

\begin{figure*}[h]
    \centering
    \includegraphics[width=\linewidth]{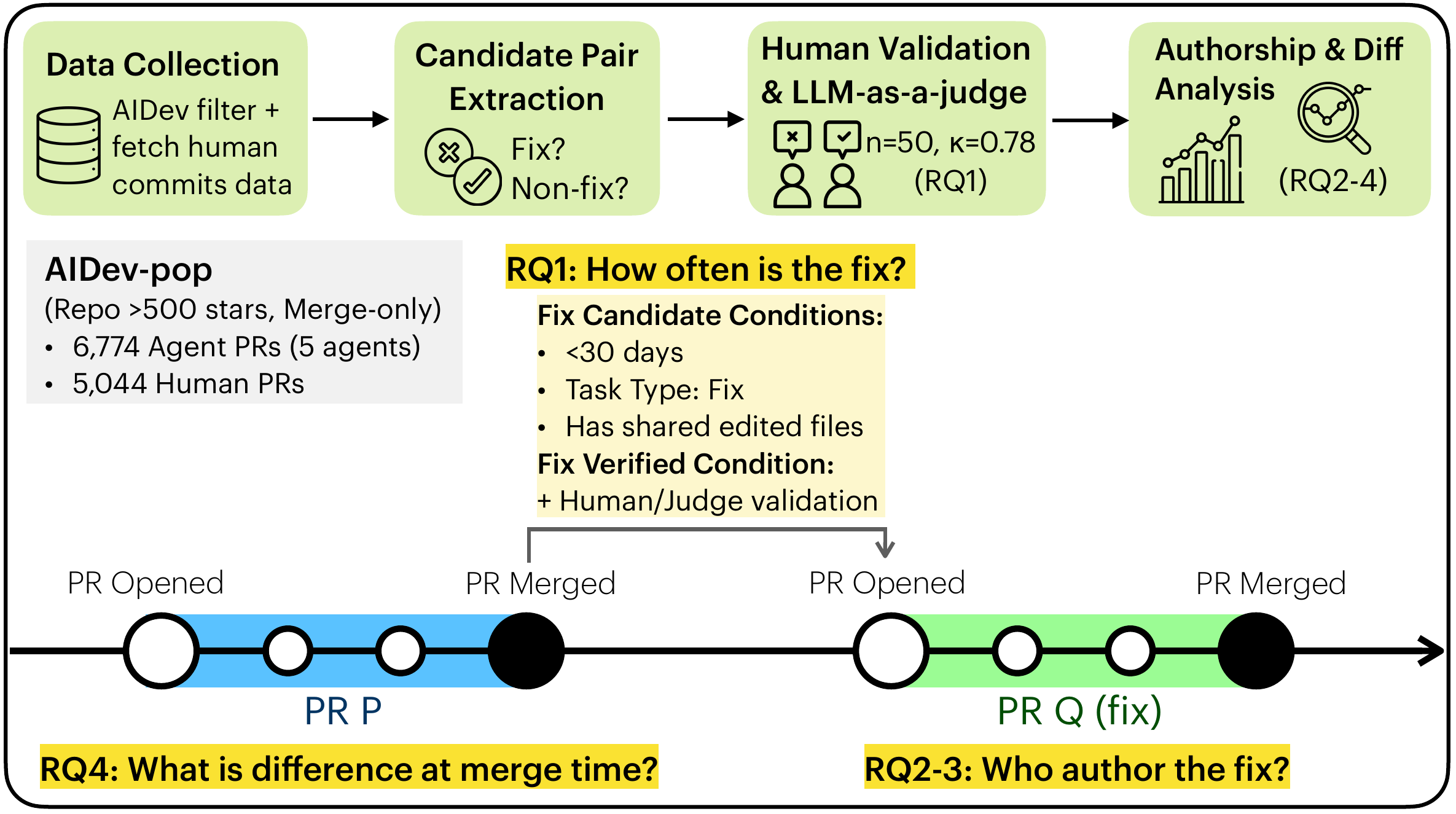}
    \caption{The overview of our study design.}
    \label{fig:study-design}
\end{figure*}

\subsection{Overview}
\label{sec:methodology:overview}

Figure~\ref{fig:study-design} shows the overview of our study.
Using merged agent PRs and merged human PRs as a baseline in the same repositories, we find the follow-up fixes in RQ1.
Given PR \emph{P} as a merged PR and PR \emph{Q} as a follow-up fix PR, we find the relationship between \emph{P$\xrightarrow{}$Q} by defining \emph{candidate} and \emph{verified} fix. 
First, we find the \emph{candidate} fixes according to the heuristic criteria. 
Then, we perform human validation and LLM judge to confirm the pairs as \emph{verified} fixes.
The verified fixes feed RQ2, which classifies who \emph{opened} each fixing PR (the same agent, another agent, or a human).
RQ3 zooms from PRs into commits which the author-class classifier labels every commit.
Finally, RQ4 compares the merges that received a follow-up fix against those that did not, on four signals visible at merge time.
Each RQ's detailed approach is described with its results in
Section~\ref{sec:results}.

\subsection{Dataset}
\label{sec:methodology:datasets}


\textbf{\aidev \citep{li2025aidev,li2025aiteammates_se3}} is a large GitHub pull request dataset of AI coding agents with more than 450,000 pull requests and 60,000 open-source repositories.
In this paper, we use \aidev-pop, a filter cohort of popular repositories in \aidev.
We follow the original filter of \aidev-pop \citep{li2025aiteammates_se3}, filtering popular repositories with more than 500 GitHub stars as of June 22, 2025.
Then, we restricted our filter to only merged PRs.

Our data yields 6{,}774 merged agent PRs across 891 repositories, covering five agents: OpenAI Codex (2{,}834), Devin (1{,}813), GitHub Copilot (1{,}428), Cursor (569), and Claude Code (130).
The agent PRs span December 24, 2024, to July 30, 2025.
The human PRs are \aidev's curated
\texttt{human\_pull\_request} table with the same $\geq\!500$-star repositories and merge-only conditions, resulting in 5{,}044 human PRs spanning January 1 to June 28, 2025.
The original \aidev's human PRs carry PR-level columns only; therefore, we re-fetch their commits, diffs, reviews, and comments per PR via the GitHub REST API.
In the cross-cohort comparison, we use the human cohort's end date as the shared cut-off for the same observation window.
All analyses are restricted to merged PRs unless otherwise stated.



\section{Results}
\label{sec:results}


\subsection{RQ1: \rqoneq}
\label{sec:results:rq1}


\subsubsection{Approach of RQ1}
\label{sec:methodology:linker}

\begin{figure}[h]
    \centering
    \includegraphics[width=\linewidth]{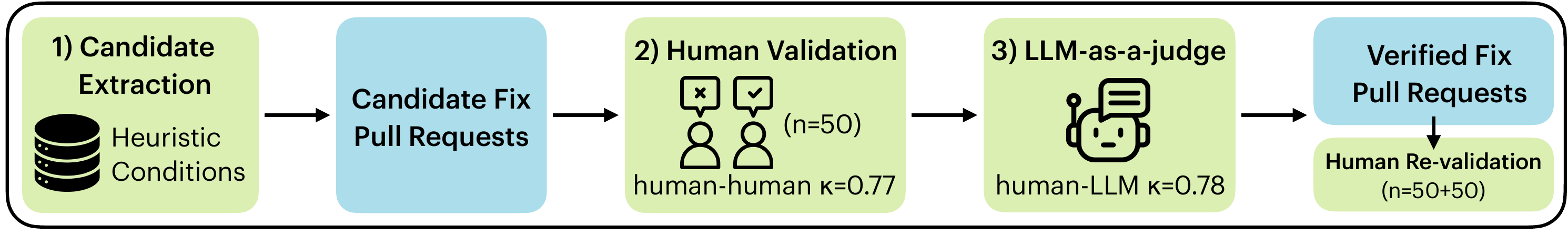}
    \caption{(RQ1) The overview of the steps to identify verified fix PRs. A green box is a process; a blue box is a PR cohort.}
    \label{fig:rq1-approach}
\end{figure}

We identify verified fix PRs in three main steps: 1) candidate extraction, 2) human validation, and 3) LLM-as-a-judge.
The overview is shown in Figure~\ref{fig:rq1-approach}.

\paragraph{\textbf{1) Candidate Extraction.}} 
Given a merged PR $P$, our goal is to find a later \emph{candidate} $Q$ that fixes PR $P$.
To do so, we apply four heuristic filters in order on both agent $P$ and human $P$ cohorts:
\begin{enumerate}
  \item $Q$ merges in the same repository, strictly after $P$, within 30 days.
  \item $Q$ is classified as \texttt{fix} type by the task-type tag in the \aidev dataset. 
  \item $Q$ co-edits at least one file that $P$ merged, identified by the intersection of net base-to-merge diff in both $P$ and $Q$.
  \item $Q$ co-edited files include at least one non-boilerplate file. This is because files such as \texttt{lockfiles}, \texttt{changelogs}, and generated artifacts churn in nearly every PR.
\end{enumerate}
Candidate fixers $Q$ are drawn from both the agent and the human cohorts, as an issue could also be fixed by a human or an agent.
Additionally, since $Q$ requires a 30-day observation period, we drop every partially observed merge $P$.
Specifically, $P$ is excluded if $P$ merged less than 30 days before the dataset end date. 
The complete details can be found in the replication package.

\paragraph{\textbf{2) Human Validation}}
We perform a human validation on the \emph{candidate} fix pairs of agent PRs to identify the verified fixes.
Two annotators (the first and second author with 3-4 years of coding experience) independently label a 50-pair sample through a web labeler (see Appendix~\ref{app:rq1-labeler}).
The PR information shown on a label page is: titles, descriptions, commit messages, and shared-file net diffs.
Each candidate pair is labeled as \emph{Direct fix} ($Q$ fixes the specific change $P$ introduced), \emph{Related touch}, or \emph{Unrelated}.
We consider a pair as a \emph{verified fix} only when labeled \emph{Direct fix}. 
On the binary verified-fix condition (\emph{Direct fix} vs. \emph{non-fix}), the two human annotators agree on 45 of the 50 sample pairs (90\,\% agreement, Cohen's $\kappa=0.77$ indicating substantial agreement~\citep{cohen1960kappa, landis1977kappa}).

\paragraph{\textbf{3) LLM-as-a-judge}}
An LLM judge (Claude Opus~4.8, temperature=0) receives the same PR information as human annotators to label the full candidate population: 2{,}510 agent-merge pairs and 2{,}206 human-merge pairs (the 50 human-validated pairs are excluded).
The LLM judge is developed to output a binary label (\emph{Direct fix} vs. \emph{Non-fix}) using the 45 agreed pairs during human validation.
After the labeling process, the judge's result is re-validated with the second author relabeling, over newly sampled pairs of agent PRs (n=50) and human PRs (n=50).
Eventually, the LLM judge substantially matches the human at $\kappa=0.78$ in agent pairs, and its \emph{Direct-fix} precision is $27/30 = 90\,\%$ on both cohorts (pooled $54/60 = 90\,\%$, Wilson 95\,\% CI = 80--95\,\%).


\subsubsection{Results of RQ1}
\label{sec:results:rq1-candidates}

\begin{table}[h]
  \centering
  \footnotesize
  \caption{(RQ1) The candidate, verified, and precision-adjusted rates on a full agent cohort.}
  \label{tab:rq1-verified}
  \setlength{\tabcolsep}{5pt}
  \begin{tabular}{lrrrr}
    \toprule
    \textbf{PR $P$ Author} &
    \shortstack[r]{\textbf{PRs}\\\textbf{($n$)}} &
      \shortstack[r]{\textbf{Candidate}\\\textbf{(\%)}} &
      \shortstack[r]{\textbf{Verified}\\\textbf{(\%)}} &
      \shortstack[r]{\textbf{Adjusted}\\\textbf{(\%)}} \\
    \midrule
    OpenAI Codex   & 2{,}049 & 24.8 & 5.5 & 4.9 (4.4--5.2) \\
    Cursor         &    153  & 15.7 & 5.2 & 5.0 (4.8--5.1) \\
    Devin          & 1{,}519 & 21.5 & 3.7 & 3.4 (3.0--3.5) \\
    Claude Code    &     63  &  4.8 & 3.2 & 3.2 \\
    GitHub Copilot &    721  & 23.4 & 3.5 & 3.1 (2.8--3.3) \\
    \midrule
    AI Agent Overall        & 4{,}505 & 22.9 & 4.5 & 4.1 (3.7--4.3) \\
    \bottomrule
  \end{tabular}
\end{table}

\textbf{Approximately 4\,\% of agent merges are followed by a verified fix.}
Table~\ref{tab:rq1-verified} shows the percentages of agent PRs that are followed by candidate or verified fix, including the precision adjusted rate. 
Among the 4{,}505 merged agent PRs with a fully observed 30-day window, $22.9\,\%$ are followed by a candidate fix ($n$=$1,032$) and $4.5\,\%$ are verified fix ($n$=$203$).
The rate ranges from $3.2\,\%$ (Claude Code, $n$=$63$) to $5.5\,\%$ (OpenAI Codex, $n$=$2,049$).
It is worth noting that the population of Claude Code is small due to half of the PRs (67 of 130) being merged after the 30-day cut-off date, so they are dropped.
Specifically, 203 of the 4{,}505 PRs carry a verified fix, including 189 judge-verified and 14 human-verified consensus on the gold pairs.


\begin{table}[h]
  \centering
  \footnotesize
  \caption{(RQ1) The cross-cohort comparison on the shared observation window (both cohorts cut at June 28, 2025). 
    }
  \label{tab:rq1-crosscohort}
  \setlength{\tabcolsep}{5pt}
  \begin{tabular}{lrrrr}
    \toprule
    \textbf{Cohort} & 
    \shortstack[r]{\textbf{PRs}\\\textbf{($n$)}} &
      \shortstack[r]{\textbf{Candidate}\\\textbf{(\%)}} &
      \shortstack[r]{\textbf{Verified}\\\textbf{(\%)}} &
      \shortstack[r]{\textbf{Adjusted}\\\textbf{(\%)}} \\
    \midrule
    AI agents & 2{,}012 & 20.9 & 3.68 & 3.34 (3.00--3.52) \\
    Humans    & 4{,}063 & 14.7 & 2.34 & 2.10 (1.87--2.23) \\
    \midrule
    \shortstack[l]{Odds ratio, within\\repositories (218 strata)} & &
      \shortstack[r]{$1.41$\\{\scriptsize[1.19--1.68]}} &
      \shortstack[r]{$1.62$\\{\scriptsize[1.10--2.39]}} & --- \\
    \bottomrule
  \end{tabular}
\end{table}

\textbf{Agent merges attract verified fixes at 1.62 times the human odds.}
In the shared observation window, $3.68\,\%$ of agent merges (74 of 2{,}012) attract a verified fix against $2.34\,\%$ of human merges (95 of 4{,}063).
Compared only within the same repositories, the gap holds: the Mantel--Haenszel odds ratio~\citep{mantel1959statistical}, stratified over the 218 repositories where both cohorts appear, is $1.62$ (95\,\% CI=$1.10$--$2.39$, $p$=$0.015$).
The Mantel--Haenzel risk ratio~\citep{greenland1985sparse} is also $1.59$.
Therefore, over the same period of time, an agent merge is roughly $1.6$ times as likely as a human merge in the same repository to be followed by a verified fix.

\begin{figure*}[h]
  \centering
  \includegraphics[width=\textwidth]{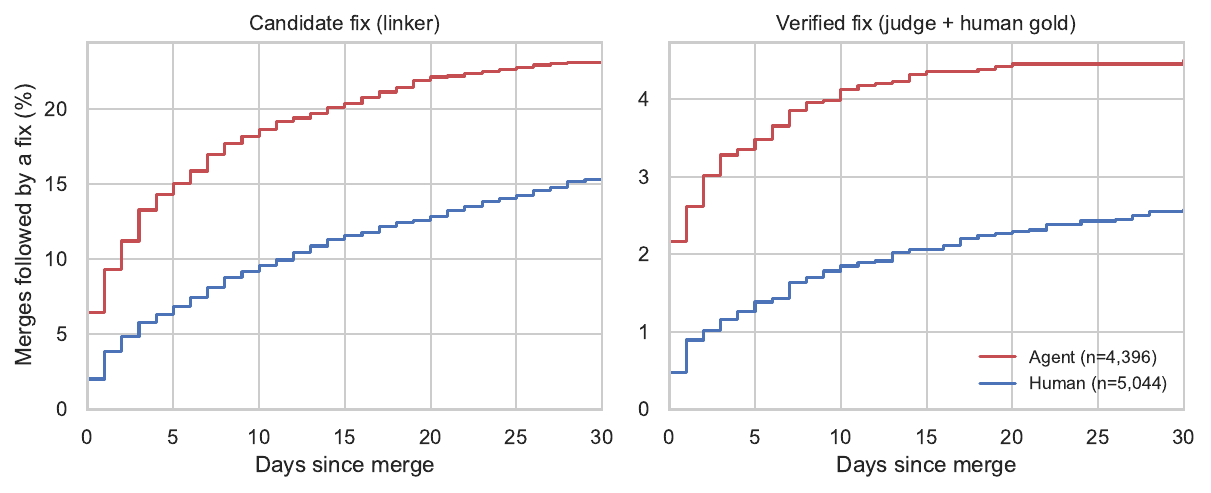}
  \caption{(RQ1) Cumulative incidence of the first follow-up fix after the merge, agent vs.\ human cohorts on the shared observation window (every merge enters with its own time).
  }
  \label{fig:repair-survival}
\end{figure*}

\textbf{The fixes arrive early, and the gap is present from the first day.}
We report Kaplan-Meier cumulative incidence curves~\citep{kaplan1958nonparametric} in Figure~\ref{fig:repair-survival}.
The time-to-event view shows every merge up to the shared cut-off with its own observation time.
The curves are separate from the first day, showing that agent PRs receive more fixes than human PRs under both the candidate and verified fix definitions.
In fact, half of the 30-day incidence accumulates within the first week in both agent and human cohorts, and the cumulative incidence reaches $4.5\,\%$ vs.\ $2.6\,\%$ by day 30 on the verified fix.

\begin{rqanswer}{RQ1}
\rqonea
\end{rqanswer}

\subsection{RQ2: \rqtwoq}
\label{sec:results:rq2}

In this research question, our goal is to track the author of the verified fix (PR $Q$), whether it is from the agent itself, another agent, or a human. 

\subsubsection{Approach of RQ2}
\label{sec:methodology:fixer}

For all \emph{verified} pairs, we classify the author by the identity that opened the fixing PR $Q$.
The author is classified into three types: \texttt{human}, \texttt{self-agent}, and \texttt{other-agent} by the following criteria:
\begin{itemize}[label=$\cdot$]
    \item \texttt{human}: when $Q$ comes from human 
    \item \texttt{self-agent}: when $Q$ is an agent PR opened by the same agent as $P$
    \item \texttt{other-agent}: otherwise, including bots
\end{itemize}
We report the split over verified fixes. 
The same classification is applied to the verified pairs of human cohort as a baseline on who fixes human merges.

\subsubsection{Results of RQ2}

\begin{table}[h]
  \centering
  \footnotesize
      \caption{(RQ2) The author of the verified fix, i.e., PR $Q$.}

  \label{tab:rq2-fixer}
  \setlength{\tabcolsep}{4pt}
  \begin{tabular}{lrrrr}
    \toprule
    \textbf{PR $P$ Author} & \textbf{Pair ($n$)} & \textbf{Self-agent (\%)} &
      \textbf{Human (\%)} & \textbf{Other agent (\%)} \\
    \midrule
    OpenAI Codex   & 125 & 59.2 & 38.4 & 2.4 \\
    Devin          &  88 & 72.7 & 23.9 & 3.4 \\
    GitHub Copilot &  40 & 95.0 &  5.0 & 0.0 \\
    Cursor         &   8 & 87.5 &  0.0 & 12.5 \\
    Claude Code    &   2 &  0.0 & 50.0 & 50.0 \\
    \midrule
    AI Agent Overall & 263 & 69.6 & 27.4 & 3.0 \\
    \midrule
    Human            & 110 & ---  & 89.1 & 10.9 \\
    \bottomrule
  \end{tabular}
\end{table}

\textbf{Agents largely fix their own merges by 69.6\,\%.}
Table~\ref{tab:rq2-fixer} shows the authors of verified fix (PR $Q$). 
Classifying the fixer of every verified agent-merge pair by who opened the fixing PR $Q$: $69.6\,\%$ are \emph{self}-fixes, meaning that the same agent that shipped the merge also ships the fix, against $27.4\,\%$ fixed by a human and $3.0\,\%$ by a different agent ($n$=$263$ verified fixes).
The pattern holds for every agent with a non-trivial count.
Codex is the agent whose merges humans most often fix (38.4\,\%), while Copilot self-fixes about $95\,\%$ of its merges.
Note that Claude Code's population is small from RQ1, because most PRs fall after the 30-day cut-off date. 

\textbf{Human merges are also fixed by humans for 89.1\,\%.}
The baseline analysis of the verified fixes of the human cohort (Table~\ref{tab:rq2-fixer}, Human row) also shows that $89.1\,\%$ are fixed by humans and $10.9\,\%$ by an agent ($n$=$110$).
Fixing work thus stays largely within the authoring population on both sides of the comparison.
Agents fix their own merges, and humans fix theirs.

\begin{rqanswer}{RQ2}
\rqtwoa
\end{rqanswer}

\subsection{RQ3: \rqthreeq}
\label{sec:results:rq3}

In this research question, we zoom into the commit-level of PR $Q$ to answer who actually authors those commits in the verified fix PRs. 

\subsubsection{Approach of RQ3}
\label{sec:methodology:author-class}

We classify every commit on each merged PR as \texttt{agent}, \texttt{automation}, or \texttt{human}, extending the rule-based
comment-classifier of \citet{duma2026reviews} (96.5\,\% accuracy on their 800-comment validation sample).
Specifically, we apply two main rules: 
\begin{enumerate} 
    \item The \textit{author login} matches agent or bot account patterns (see Appendix~\ref{app:rq3-approach}).
    \item The \textit{commit message} matches agent-generated message (see Appendix~\ref{app:rq3-approach}).
\end{enumerate}
Note that OpenAI Codex does not sign its commits (only PR-level). 
In fact, they are authored under the developer's GitHub login, and fewer than 2\,\% carry any marker.
Therefore, Codex's split is excluded from the commit measure and we report only the overall commit volume.
It is also worth noting that because an agent commit that lacks both marker and agent login is counted as \texttt{human}, the agent-authored share in this work can be considered a conservative \emph{lower bound}.




In the analysis, we aggregate results \emph{per PR}, not over the aggregated commit stream.
Thus, each merged PR contributes its own agent-authored share, and we report two numbers: the average of those shares per PR as \textit{Avg Agent Share per PR} and the percentage of PRs whose commits are all agent-authored as \textit{PR/w All Commits by Agent}.
We also run the same classification over the verified fix of the human cohort for a baseline comparison.

\subsubsection{Results of RQ3}

\begin{table}[h]
  \centering
  \footnotesize
  \caption{(RQ3) Commit author-class inside the verified fix PRs, 
  aggregated per PR.
           }
  \label{tab:rq3-fix-commit-only}
  \setlength{\tabcolsep}{4pt}
  \begin{tabular}{lcccc}
    \toprule
      \textbf{PR $Q$ Author} & 
      \shortstack[c]     {\textbf{PRs}\\\textbf{($n$)}} &
      \shortstack[c]{\textbf{Commits}\\\textbf{median ($n$)}} &
      \shortstack[c]{\textbf{Avg Agent Share}\\\textbf{per PR (\%)}} &
      \shortstack[c]{\textbf{PR/w All Commits}\\\textbf{by Agent (\%)}} \\
    \midrule
    GitHub Copilot   &  40 & 3.0 & 86.6 & 70.0 \\
    Devin            &  59 & 1.0 & 92.0 & 84.7 \\
    Cursor           &  23 & 1.0 & 79.1 & 69.6 \\
    Claude Code      &   1 & 8.0 & --- & --- \\
    OpenAI Codex     & 103 & 1.0 & --- & --- \\
    \midrule
    All (excl.\ Codex) & 123 & 2.0 & 87.4 & 76.4 \\
    \midrule
    Human              & 124 & 2.0 & 0.8 & 0.8 \\
    \bottomrule
  \end{tabular}
\end{table}

\textbf{The commits added on a verified fix of agent PR are majority agent-authored with the average share of 87.4\,\%.}
Classifying every commit of the unique verified fix PRs on marker and login evidence (Table~\ref{tab:rq3-fix-commit-only}), the result shows that
agent-opened fix PRs are $79$--$92\,\%$ agent-authored per PR, and 76.4\,\% of those PRs are agent-authored throughout all commits.
On human-opened fix PRs, the same classifier carries almost no marked agent work ($0.8\,\%$ average per-PR share over 124 PRs and 554 commits).
Therefore, who opens the fix PR is, on marked evidence, also who writes it.

\begin{table}[h]
  \centering
  \footnotesize
  \caption{Commit author-class on \emph{all} merged \aidev-pop PRs.
           OpenAI Codex is unmeasurable (---) as it
           does not sign its commits.}
  \label{tab:rq3-all-commits}
  \setlength{\tabcolsep}{4pt}
  \begin{tabular}{lcccc}
    \toprule
    \textbf{PR Author} & 
      \shortstack[c]     {\textbf{PRs}\\\textbf{($n$)}} &
      \shortstack[c]{\textbf{Commits}\\\textbf{median ($n$)}} &
      \shortstack[c]{\textbf{Avg Agent Share}\\\textbf{per PR (\%)}} &
      \shortstack[c]{\textbf{PR/w All Commits}\\\textbf{by Agent (\%)}} \\
    \midrule
    GitHub Copilot   & 1{,}428 & 4.0 & 81.6 & 51.8 \\
    Devin            & 1{,}813 & 3.0 & 80.1 & 56.9 \\
    Cursor           &    569  & 2.0 & 72.9 & 54.1 \\
    Claude Code      &    130  & 2.5 & 57.1 & 40.0 \\
    OpenAI Codex     & 2{,}834 & 1.0 & ---  & ---  \\
    \midrule
    All (excl.\ Codex) & 3{,}940 & 3.0 & 78.8 & 54.1 \\
    \midrule
    Human & 4{,}900 & 2.0 & 0.9 & 0.6 \\
    \bottomrule
  \end{tabular}
\end{table}

\textbf{AI Agents write the fixes they open, more often than they write their merges.}
We zoom out to the author-class on all merged PRs in \aidev-pop (Table~\ref{tab:rq3-all-commits}).
Across the full 3,940 PRs that include 123 fix PRs, the average agent-authored share is 78.8\,\% overall
across four AI agents (exclude Codex), and 54.1\,\% of those PRs are all agent-authored commits.
This is below the 87.4\,\% and 76.4\,\% the same agents record on only their verified fix PRs.
Therefore, this result indicates that agents have a higher share of commits on fix PRs than on their merged PRs in general.



\begin{rqanswer}{RQ3}
\rqthreea


\end{rqanswer}

\subsection{RQ4: \rqfourq}
\label{sec:results:rq4}

In this research question, our goal is to find the characteristics that differ between agent PRs that are followed by a fix and those that are not.

\subsubsection{Approach}
\label{sec:methodology:visibility}

To compare the distributions, we use four merge-time signals of agent merges: time under review, commits per PR, post-first-push churn, and review items, shown as split violins.
For each signal we test the two-group difference with a Mann--Whitney
U test~\citep{mann1947test} and measure its magnitude with Cliff's
$\delta$~\citep{cliff1993dominance,romano2006appropriate} against the magnitude labels per \citet{romano2006appropriate}. 



Additionally, because pooled contrasts can reflect repository confounds rather than merge properties, each signal is also tested \emph{within} repositories with a conditional logistic regression~\citep{chamberlain1980conditional}, one model per signal:
\begin{equation}
  \operatorname{logit} P\!\left(Y_{ri}=1\right)
    \;=\; \alpha_r
    \;+\; \beta \log_{10}\!\left(1 + s_{ri}\right)
    \;+\; \gamma \log_{10}\!\left(1 + f_{ri}\right),
  \label{eq:rq4-clogit}
\end{equation}
where $Y_{ri}$ indicates that merge $i$ in repository $r$ is followed by a verified fix within 30 days; $s_{ri}$ is the merge-time signal;
$f_{ri}$ is the number of non-boilerplate files shipped (i.e., PR size control); and $\alpha_r$ is a repository control.
The reported odds ratio $e^{\beta}$ compares two same-repository merges ten times apart on each $\text{signal}$ due to a log-scale (Figure~\ref{fig:rq4-forest}).

\subsubsection{Results}

\begin{figure*}[h]
  \centering
  \includegraphics[width=\textwidth]{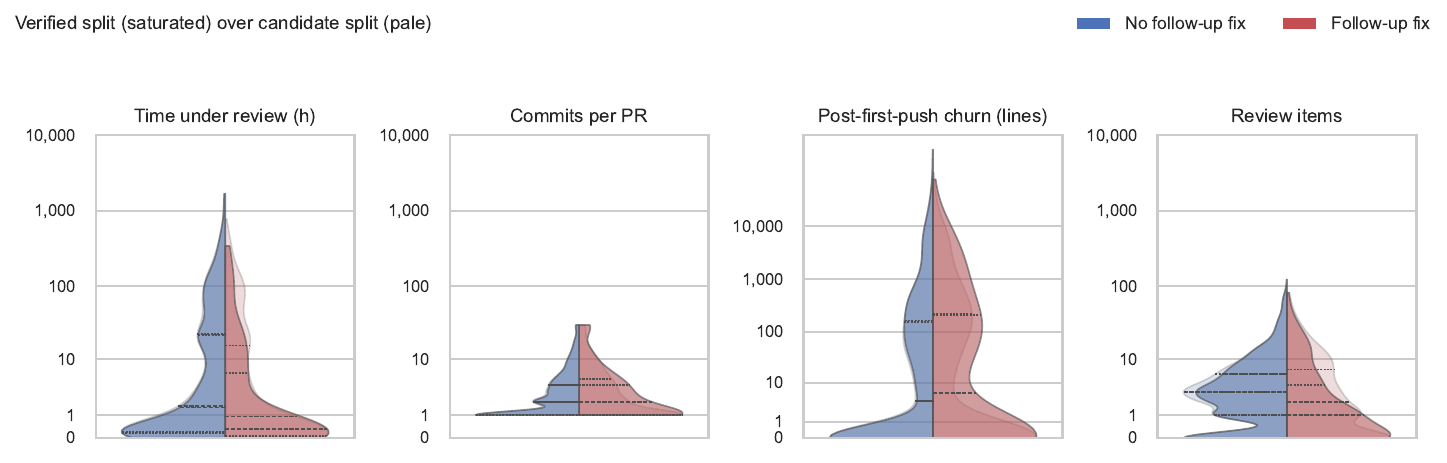}
  \caption{(RQ4) The signal distributions of agent PRs with and without a follow-up fix. 
           }
  \label{fig:violin-signals}
\end{figure*}

Figure~\ref{fig:violin-signals} shows the four signal distributions of the
two groups side by side, and Table~\ref{tab:rq4-signals} shows the significance tests and their effect sizes.

\begin{table}[h]
  \centering
  \footnotesize
  \caption{(RQ4) Per-signal comparison of agent PRs with and without a follow-up fix.
  A negative $\delta$ means the w/ follow-up fix merges sit \emph{lower} on the signal.
           }
  \label{tab:rq4-signals}
  \setlength{\tabcolsep}{4pt}
  \begin{tabular}{lrrrr}
    \toprule
    & \multicolumn{2}{c}{\textbf{Candidate}} &
      \multicolumn{2}{c}{\textbf{Verified}} \\
    \cmidrule(r){2-3}\cmidrule(l){4-5}
    \textbf{Signal} & \textbf{p} & \textbf{$\delta$ Effect size} &
      \textbf{p} & \textbf{$\delta$ Effect size} \\
    \midrule
    Time under review      & $<$0.001 & $-$0.100 (negl.) & $<$0.001 & $-$0.221 (small) \\
    Commits per PR         & 0.15     & $+$0.028 (negl.) & 0.75     & $+$0.013 (negl.) \\
    Post-first-push churn  & 0.21     & $+$0.024 (negl.) & 0.41     & $+$0.032 (negl.) \\
    Review items           & $<$0.001 & $-$0.101 (negl.) & $<$0.001 & $-$0.239 (small) \\
    \bottomrule
  \end{tabular}
\end{table}

\textbf{In the pooled comparison, the two groups are nearly identical, with the follow-up fix PRs group having slightly less time under review and fewer review items (small effects).}
Our four signals show that the difference in effect size is negligible to small (Table~\ref{tab:rq4-signals}). 
The two differences that reach the significance point are in \textit{negative} effect sizes.
Specifically, the merged agent PRs with a follow-up fix yield slightly shorter review time and fewer review items than the group without a follow-up fix.
The result indicates that agent PRs with a lower code review (time and items) correlate with slightly more follow-up fixes in the pooled view.

\begin{figure}[h]
  \centering
  \includegraphics[width=0.7\columnwidth]{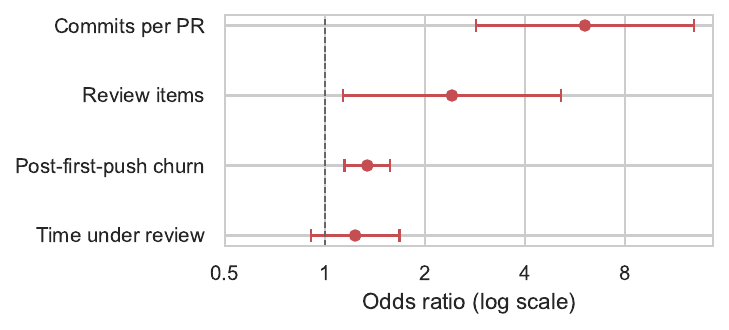}
  \caption{Within-repository, odds of a verified follow-up fix, per signal.
           Each dot is the odds ratio comparing merges ten times apart on
           a $\text{signal}$, with a 95\,\% confidence interval. 
           }
  \label{fig:rq4-forest}
\end{figure}

\textbf{In the within-repositories comparison, the effort to complete a PR could mark a follow-up fix.}
When the repository and the number of shipped files are held fixed, the result changes (Figure~\ref{fig:rq4-forest}). 
Our finding shows that when a PR has ten times the commits of the other, the one with more commits has 6.1 times the odds of being followed by a verified fix ($p<0.001$).
On the same tenfold scale, review items carry $2.4$ times ($p=0.022$), churn carry $1.3$ times ($p<0.001$), and time under review shows no significant effect ($p=0.18$).
Therefore, the result shows that the PRs that needed many rounds of work before the merge are the ones most likely to need more work after it, when compared within the repository. 


\begin{rqanswer}{RQ4}
\rqfoura
\end{rqanswer}

\section{Discussion}
\label{sec:discussion}


In this section, we discuss our findings across research questions. 

\subsection{The merge is not the end of the work, and code review is needed.}
\label{sec:discussion:endpoint}

Previous studies of agent PRs largely stop at the merge (e.g., size, review activity, and acceptance~\citep{li2025aiteammates_se3,ogenrwot2026modify, duma2026reviews}) as a finished work.
Our results show that the work continues, at a small but consistent rate: $4.5\,\%$ of agent merges carry a verified fix within 30 days, which is $1.62$ times more than human merges (RQ1).
However, our RQ4 findings also show that, at the aggregated level, the signals at the merge time between PRs with and without a follow-up fix have only minor differences. 
Therefore, the study indicates that a PR should also be monitored beyond the merge, and the signal may live in the code rather than in the merge metadata. 
In fact, on \aidev{} PRs, a content-level model (i.e., code semantics and lexicon) that reads the code changes can predict the post-merge modification at AUC=$0.671$~\citep{rahman2026survive}.
Therefore, in line with our study, this indicates that verification of agent contributions requires code review (i.e., reading the code changes).
\textit{Future research} should measure how much of the agent fix that code review can capture and explore a monitoring framework for the agent platform to catch issues before and after a merge.

\subsection{Owners fix their own code, but now the owner is an agent.}
\label{sec:discussion:ownership}

Code ownership has been a long-standing focus in software engineering for many years.
Code changes are recommended to be made by an owner rather than by outsiders for less defects~\citep{bird2011ownership,greiler2015ownership}.
Before the age of AI coding agents, humans fix roughly half of their own defects at individual granularity~\citep{zhu2021meaculpa,keller2023ossfuzz,tan2022selffixed}.
This is in line with our findings (RQ2 and RQ3) that human merges are still fixed by humans $89\,\%$ of the time.
Similarly to humans, our results indicate that agent merges are also fixed by the same agent $69.6\,\%$ of the time.
So, our findings show the same pattern of self-fixing in both agent and human cohorts, similar to the history, but with the new \textit{operational} owner.
This partially addresses the concern that the AI agent code goes unmaintained~\citep{sawada2026maintenance}, at least at the fix level.
In human teams, abandoned code is adopted by experienced insiders~\citep{rigby2016knowledgeloss}, and that adoption depends on someone having accumulated the experience while the owner was active; an agent that authors a merge and then works on its own fixes leaves no such insider behind.
Therefore, the next question is about the deeper granularity of what happens to the agent code after the agent \textit{model} that owns the merge is deprecated or discontinued, and whether projects should adopt explicit ownership policies similar to those for human.
\textit{Future research} should explore the model granularity and how the model and agent ownership affects the codebase.  

\section{Threats to Validity}
\label{sec:threats}



\paragraph{Construct Validity.}
Author-class counts measure who \emph{committed or marked} a change, not who wrote it, and OpenAI Codex's commits carry no marker. 
We mitigate this by reporting the agent-authored shares (RQ3) as floors and excluding Codex from the commit decomposition. 
The \emph{candidate fix} label links by file co-location within 30 days, so it can miss cross-file fixes, fixes slower than 30 days, and fixes folded into non-\texttt{fix}-typed PRs. 
We mitigate this by applying identical linking criteria to both cohorts, so the cross-cohort comparison holds, and by reading the verified rates as floors.
Commit fetches cap at 30 per PR (\aidev's cap); under 2\,\% of PRs reach it, so we accept the truncation.

\paragraph{Internal Validity.}
Agent labels in \aidev{} derive from commit-author patterns, so an unmarked agent PR hiding in the \emph{human} table would count as a human author. 
This error can inflate the human-fixed share; however, the results show that our findings remain the same for agents consistently fixing more than half of their code. 
Additionally, fixer identity captures the authorship side of ownership only; we acknowledge that ownership can also be held through code review~\citep{thongtanunam2016revisiting} and strictly scope the claim to only who authors the fix.


\paragraph{External Validity.}
Our population is the five \aidev{} agents in popular ($\geq$500-star) repositories, so other agents, low-activity repositories, and behaviour after the data window are out of scope; we state the findings for this population rather than for agent PRs in general. 
Our snapshot admits more repositories than \cite{li2025aidev}'s ($1{,}479$ vs.\ $856$ at $\geq$500 stars) because stars and PRs accrue over time. We verified that the snapshot reproduces their $\geq$101-star curated set exactly ($2{,}807$ repositories), so the difference is accrual, not a divergent population.


\paragraph{Conclusion Validity.}
Verification is reliable at the binary Direct-fix level but fluctuate at three-way labels (human--human $\kappa$=$0.77$ binary vs.\ $0.60$ three-way; judge--human $0.78$ vs.\ $0.46$). 
We mitigate this by using the binary construct for every headline claim. 
An LLM judge could favour one cohort, so we checked this with blind spot-checks on both sides: its Direct-fix precision is $27/30=90\,\%$ on both agent-merge and human-merge pairs.
Thus, the adjusted rates scale by the pooled estimate ($54/60=90\,\%$, Wilson 80--95\,\%). 
Pooled cross-cohort rates would mix time and repository effects. 
We mitigate this by comparing on the shared observation window with repository-stratified (Mantel--Haenszel) estimates. 
The single-cohort rates keep each cohort's own window because the shared cut-off shrinks the agent cohort unevenly (e.g., leaving Cursor with four observable merges and GitHub Copilot with no verified fix); we therefore never compare cohorts on the own-window rates.

\section{Conclusion}
\label{sec:conclusion}


A merged AI-agent pull request is usually counted as finished work, but previous studies show that agent work can leave issues behind.
In this paper, we ask how often an agent PR is followed by a fix and who actually finishes it.
We tracked 6{,}774 merged agent PRs in \aidev-pop into their follow-up fixes, compared to a human baseline in the same repositories.
We found that agent merges attracted verified fixes at $1.62$ times the odds of human merges over the same observation window.
Of the verified fixes in agent merges, $69.6\,\%$ come from the same agent.
The commits to verified fix PRs of agent merges were substantially agent-authored (on average $87.4\,\%$ per PR).
The merges that later needed a fix have nearly the same signals as the others at merge time; the clearest signal, within repositories, was how many commits the merge itself had taken.
Agents largely finish their own job on both sides of the merge, but their merges need fixing modestly more often than human merges, with little visibility up front.
As agents author more of the code that ships, following their work past the merge is what will keep evaluation aligned with how software is now produced.




\section*{Statements and Declarations}

\paragraph{Funding:} This work is supported by the Australian Composites Manufacturing CRC under Grant Number: HPN003 -- ForgeX.ai, AI factory management (Phase 1).

\paragraph{Author Contributions:} 

The conception and design of the study, data collection and data analysis were carried out by Wannita Takerngsaksiri. 
Materials preparation and human validation were performed by Wannita Takerngsaksiri and Nhat Duong.
The first draft of the manuscript was written by Wannita Takerngsaksiri.
The review and acquisition of funding was done by Scott Barnett.
All authors read, provided comments, and approved the final manuscript.

\paragraph{Data Availability Statement:}The replication package is available at \url{https://github.com/wannita901/pr-fix-authorship}. 
The pull requests are drawn from the public AIDev dataset, mined from public GitHub repositories; no private or personally identifiable data are used.

\paragraph{Conflict of Interest:} The author declares no conflict of interest.

\paragraph{Ethical approval:} Not applicable.

\paragraph{Informed consent:} Not applicable.

\paragraph{Clinical Trial Number:} Not applicable.





\bibliographystyle{spbasic}
\bibliography{references}


\appendix

\sloppy

\section{The Labeler for Verified Fix Annotation}
\label{app:rq1-labeler}

\begin{figure}[h]
    \centering
    \includegraphics[width=\linewidth]{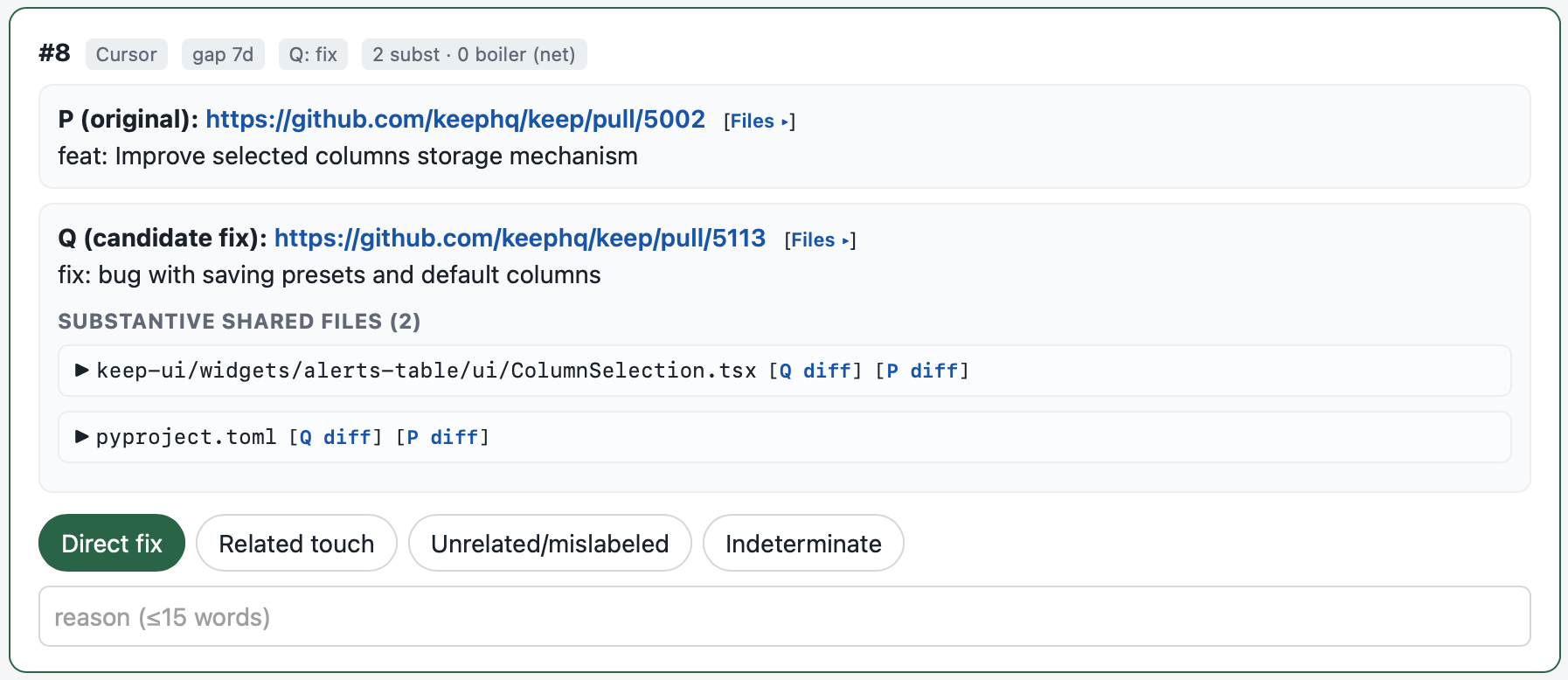}
    \caption{A sample screen of web labeler for human to annotate a verified fix.}
    \label{fig:rq1-web-labeler}
\end{figure}

Figure~\ref{fig:rq1-web-labeler} shows a web labeler screen that human annotators use to label the candidate pair: Original $P$ $\xrightarrow{}$ Candidate Fix $Q$ (\S\ref{sec:methodology:linker}). 
Each pair is labeled with one of the four labels defined below:
\begin{itemize}[label=$\cdot$]
  \item \emph{Direct fix}: $Q$ exists to \emph{repair} a defect, regression, or wrong/incomplete behavior that $P$ introduced or left behind; $Q$'s modifies or reverts the lines $P$ added in the shared files.
  \item \emph{Related touch}: $Q$ does \emph{incremental or additive work} in the same
    area as $P$ (e.g., extending or refactoring it) without repairing a defect.
  \item \emph{Unrelated/mislabeled}: $Q$ touches the shared file for an \emph{independent reason}.
  \item \emph{Indeterminate}: the material is \emph{insufficient} to decide.
\end{itemize}
We consider a pair a \emph{verified fix} only when labeled \emph{Direct fix}.

\section{The Rule List of Commit Author-Class Mapping (RQ3)}
\label{app:rq3-approach}

There are two main rules to map the evidence of author-class in RQ3 (\S\ref{sec:methodology:author-class}): \textit{author login} and \textit{commit message}.
Table~\ref{tab:rq3-approach} is the complete set of commit author-class rules.

\begin{table}[h]
  \centering
  \footnotesize
  \caption{The author-class evidence rules per agent.
           }
  \label{tab:rq3-approach}
  \setlength{\tabcolsep}{3pt}
  \resizebox{\columnwidth}{!}{%
  \begin{tabular}{@{}p{0.17\linewidth}p{0.4\linewidth}p{0.5\linewidth}@{}}
    \toprule
    \textbf{Class} & \textbf{Author Login} & \textbf{Commit Message} \\
    \midrule
    GitHub Copilot & \texttt{Copilot} & \texttt{Co-Authored-By: Copilot} \\
    \addlinespace
    Devin & \texttt{devin-ai-integration[bot]} & \texttt{Devin Run Link};
      \texttt{Co-Authored-By: Devin} \\
    \addlinespace
    Cursor & \texttt{cursoragent} & \texttt{Co-Authored-By: Cursor} \\
    \addlinespace
    Claude Code & \texttt{claude[bot]}, \texttt{claude-code} &
      \texttt{Generated with} $+$ robot emoji; \texttt{Generated with [Claude Code}; \texttt{Co-Authored-By: Claude} \\
    \addlinespace
    OpenAI Codex & \texttt{codex}, \texttt{codex-cli} 
    & \texttt{Generated
      with \ldots{} Codex}; \texttt{Co-Authored-By: openai-codex} 
      \\
    \midrule
    automation & any other \texttt{*[bot]}; \texttt{dependabot},
      \texttt{renovate}, \texttt{codecov}, \texttt{github-actions},
      \texttt{autofix-ci}, \texttt{vercel}, \texttt{netlify},
      \texttt{semantic-release-bot}, \texttt{pre-commit-ci},
      \texttt{allcontributors} & --- \\
    \midrule
    human & everything else (default) & --- \\
    \bottomrule
  \end{tabular}
  }
\end{table}

\end{document}